\documentstyle[prd,aps,floats,preprint,epsfig,12pt]{revtex}
\tighten
\begin{document}
\draft
\date{\today}
\preprint{\vbox{\baselineskip=13pt
\rightline{LPTHE Orsay-99/07}
\vskip0.2truecm
\rightline{hep-ph/9901400}}}
\title{LOCAL COSMIC STRINGS FROM PSEUDO-ANOMALOUS U(1).}
\author{C\'edric Deffayet}
\address{LPTHE, Universit\'e Paris-XI, B\^atiment 211, F-91405 Orsay Cedex, France}
\maketitle
\begin{center}Peyresq Physics III, Proceedings, 1998 \end{center}
\begin{abstract}
Local cosmic strings solutions are introduced in a model with a peudo-anomalous U(1) gauge 
symmetry. Such a symmetry is present in many superstring
compactification models. The coupling of those strings with the axion
necessary in order to cancel the anomalies does not prevent them from
being local, even though their energy per unit length is found to
diverge logarithmically. We discuss briefly the formation of such strings and
the phenomenological constraints that apply to their parameters.
\end{abstract}
\section*{Introduction}
This talk is based on 
a work done in collaboration with P.Bin\'etruy  and P.Peter  \cite{nous} about cosmic strings in models with a pseudo-anomalous
$U(1)$ symmetry. Such a symmetry arises generically in a large class of superstring
 compactification models as
a remnant of the Green-Schwarz~\cite{GS} mechanism of anomaly cancellation in the underlying 10
dimensional supergravity. In a bottom top approach, interest for such model has recently been
renewed in the framework of horizontal symmetries trying to explain the hierachies in the quarks
and leptons spectrum \cite{hierar}.

 It is generically well known that cosmic strings may form in the early universe
 in the breaking of a $U(1)$ symmetry \cite{kibble,NO}; this is also true when the symmetry is
 pseudo-anomalous \cite{casas,harvey,jeannerot,LR}. However because of their being coupled to the 
 axion field, such strings were thought to be of the global kind. We show that there exists 
 a possibility that (at least some of) the strings formed at the breaking of this anomalous U(1)
 be local, in the sense that their energy per unit length can be localized in a finite region 
 surrounding the string core, even though this
 energy is formally logarithmically infinite. It will be shown indeed
 that the axion field configuration can be made to wind around the
 strings so that any divergence must come from the region near the core
 instead of asymptotically. The cutoff scale that must then  be
 introduced is thus a purely local quantity, definable in terms of the
 microscopic underlying fields and parameters. It is arguable that
 such a cutoff should be interpreted as the scale at which the
 effective model used throughout ceases to be valid.
 
 This talk is divided into three parts, in the first one I expose the model which 
 we use for studying string solutions. Then I talk
 briefly about the Higgs mechanism in this framework; and in a third part I specifically give
 the local cosmic strings solutions and discuss some related phenomenological questions.   

\section{The model and its raison d'etre}

Compactification models for the heterotic string 
are known to lead in general to the presence in the 4-dimensional
 theory of a so-called universal axion field $a$ \cite{U1}. At the supersymmetric level,
  this pseudoscalar field belongs to the same (chiral)
supermultiplet $S$ as the dilaton field $s$ and they form a complex scalar
field $s+ia$.  The superfield $S$ couples in a model independent way to
the gauge fields present in the theory; one has in particular in the lagrangian:
\begin{equation}
{\cal L} = -{s \over 4 M_{_P}} \sum_a F^a_{\mu\nu} F^a_{\mu\nu}
           +{a \over 4 M_{_P}} \sum_a F^a_{\mu\nu} \widetilde
F^a_{\mu\nu},
\label{debut}
\end{equation}
where $F^a_{\mu\nu}$ is the field strength associated with the gauge field  
$A_\mu^a$ and $M_{_P}$ is the reduced Planck scale, the index $a$ runs over
all gauge groups and
\begin{equation} \widetilde F^{a\mu\nu}\equiv {1\over 2}
\varepsilon^{\mu\nu\rho\sigma}
F^a_{\rho\sigma}.\label{deftilde}
\end{equation} 
An abelian symmetry with gauge field $A_\mu$ may have (mixed)
anomalies: under a gauge transformation of parameter $\alpha$, the effective lagrangian is no
longer invariant but picks up new terms (the anomaly) given by:
$\delta {\cal L} = -{1 \over 2} \delta_{_{GS}} \alpha \sum_a
F^a_{\mu\nu}
\widetilde F^a_{\mu\nu}.$
This can be cancelled by an apropriate shift of the axion
$a$.  Since there is a single model-independent axion (in the weakly coupled heterotic string 
spectrum), only one
abelian symmetry, henceforth referred to as $U(1)_X$, may be
pseudo-anomalous. 

One can write the supersymmetric lagrangian of a model with such a pseudo-anomalous $U(1)_X$ symmetry:

\begin{equation}
{\cal L} = \left({\cal K} + A_{i}^\dagger
e^{X_iV} A_{i} \right)_{D-term} + 
\left(\frac{1}{4}SW^{\alpha}W_{\alpha}\right)_{F-term} + h.c. \label{susyl}
\end{equation}
Using the standard notations of \cite{wess}. $S$ is the axion-dilaton superfield, $V$ is the gauge vector
superfield, and $A_i$ are chiral superfields of respective charge $X_i$ under the pseudo
anomalous $U(1)_X$ symmetry. , and ${\cal K}= -\ln (S + \bar S -
4\delta_{_{GS}} V)$ is the modified K\"ahler function for S \cite{U1}.
The D-term of the lagrangian is now invariant under the following transformations:

\begin{eqnarray}
A_{i} \rightarrow e^{-q_{i}\Lambda}A_i, \\
V \rightarrow V + i(\Lambda - \Lambda^\dagger),\\
S \rightarrow S + 4i \delta_{GS}\Lambda.
\end{eqnarray}
Where $\Lambda$ is a chiral superfield which is the generalized gauge transformation parameter.
The variation of the term $SW^{\alpha}W_{\alpha}$ under a restricted gauge transformation
 compensates for 
the 1-loop appearance of the gauge anomaly. A superpotential ${\cal W}(A_i)$ can also be added, one can show
that it receives no contribution from $S$ in perturbation theory (however it may no longer be the case at
the non perturbative level).
Integrating out the auxiliary fields, one obtains for the bosonic terms in the lagrangian:

\begin{eqnarray}
{\cal L} = -\frac{M^2_{_P}}{4s^2} \partial^{\mu}s \partial_{\mu}s - (D_{\mu}\Phi_i )^\dagger (D^{\mu}\Phi_i)
-\frac{1}{4}\frac{s}{M_{_P}} F_{\mu\nu} F^{\mu\nu} \nonumber \\
+\frac{1}{4} \frac{a}{M_{P}} F_{\mu\nu} 
\widetilde {F}^{\mu\nu} - \frac{M^4_{_P}}{s^2} \left(\frac{\partial ^\mu a}{2M_{_P}} - \delta_{GS} A^{\mu}\right)
^2 -\frac{M_{_P}}{2s} \left(\frac{ \delta_{GS}M^3_{_P}}{s} + q_i \Phi_i^\dagger \Phi_i \right)^2.
\label{susypot}
\end{eqnarray}
Where we have restored the Planck mass everywhere and we have introduced 
the scalar fields $\Phi_i$ carrying the integer charge $X_i$ ($X_i$ 
has been rescaled by a factor $2$) under the $U(1)_X$ symmetry which are the lowest component
 of the chiral superfield $A_{i}$.
  The covariant
derivative is defined by
\begin{equation}
D^{\mu}\Phi_i \equiv (\partial^\mu - iX_i A^{\mu})\Phi_i
\label{defderiv}.\end{equation}

The $\delta_{GS}$ parameter (which fixes 
the scale of the symmetry breaking with respect to the
fundamental scale of the theory, here given by $M_{_P}$), may be computed in the 
framework of the weakly coupled string 
and is found to be~\cite{U1}:

\begin{equation}
\delta_{_{GS}} = {1 \over 192 \pi^2} \sum_i X_i, \label{weaks}
\end{equation}
where $X_i$ are the charges of the different fields under $U(1)_X$.

We then obtain the model which we considered \cite{nous} by  setting the dilaton to its v.e.v
$\langle s\rangle ={M_{_P}}/{g^2}$ which gives the gauge coupling constant of the $U(1)$
symmetry. Rescaling the Green
Schwarz coefficient $\delta_{GS}$ and the axion by a factor $g^{2}$, one finds:

\begin{eqnarray} {\cal L} &=& 
-(D_{\mu}\Phi_i )^\dagger (D^{\mu}\Phi_i)\nonumber\\
& & -{1\over 4g^2} \left( F_{\mu\nu} F^{\mu\nu} -{a\over
M_{_P}}F^{\mu\nu}
\widetilde F_{\mu\nu}\right)\nonumber \\
& & -\delta_{_{GS}}^2  M_{_P}^2 A_\mu A^\mu +
\delta_{_{GS}}  M_{_P} A^\mu \partial _\mu a \nonumber\\
& & -{1\over 4} \partial _\mu a \partial^\mu a -V(\Phi_i) .\label{lag}\end{eqnarray} 
The potential $V(\Phi_i)$ is defined  by
\begin{equation}
V(\Phi_i) \equiv {g^2\over 2} (\Phi_i^\dagger X_i \Phi_i +\delta_{_{GS}}
M_{_P}^2)^2 .\label{pot}
\end{equation}
The lagrangian ~(\ref{lag}) is now invariant under the following local
gauge transformation with gauge parameter $\alpha(x^\mu)$
\begin {eqnarray} \Phi_i &\rightarrow& \Phi_i e^{iX_i\alpha}, \nonumber
\\
A_{\mu} &\rightarrow& A_{\mu} + \partial_{\mu}\alpha, \label{gauge}\\
a&\rightarrow&a+2M_{_P}\delta_{_{GS}}\alpha. \nonumber \end{eqnarray}
The transformation of the term $(a/ 4g^2 M_{_P})F^{\mu\nu}
\widetilde F_{\mu\nu}$ cancels the variation of the effective 
lagrangian due to the anomaly, namely $\delta {\cal L} = -(1/2g^2
)\delta_{_{GS}}\alpha F^{\mu\nu} \widetilde F_{\mu\nu}$ (assuming we
are also transforming the fermions of the theory not written
explicitely in ~(\ref{lag})).  Making a rigid gauge transformation
with parameter $\alpha = 2\pi$ without changing $a$ as a first step
but transforming the other fields (including the fermions), leads us
to interpret $a$ as a periodic field of period $4\pi
\delta_{_{GS}}M_{_P}$ through the redefinition $a
\rightarrow a-4 \pi \delta_{_{GS}}M_{_P}$ which leaves
the lagrangian invariant. It is also manifest that $a$ behaves like a phase, in the following
rewriting of the kinetic term and of the axionic $\theta$-term in
${\cal L}$:

\begin {eqnarray}
{\cal L}_{kin,\theta}&=&-\frac{1}{4g^2}\left(F^{\mu\nu}F_{\mu\nu} -
\frac{a}{M_{_P}}F^{\mu\nu}\widetilde F_{\mu\nu}\right) \nonumber\\
& &  -\partial^{\mu}\phi_i\partial_{\mu}\phi_i -
{\cal D}^{\mu}\eta_i {\cal D}_\mu \eta_i 
- {\cal D}^{\mu}a {\cal D}_\mu a
\end {eqnarray}
where we have defined $\Phi_i\equiv\phi_{i}e^{i\eta_i}$ ($\phi_i$ being
the modulus of $\Phi_i$) and:
\begin{eqnarray} 
{\cal D}^{\mu}\eta_i =\phi_i X_i \left(\frac{\partial^{\mu}\eta_i}{X_i}-A^{\mu}\right),
\\
 {\cal D}^{\mu}a =  {M_{_P}}\delta_{_{GS}}\left(\frac{\partial^{\mu}a}
{2M_{_P}\delta_{_{GS}}}-A^{\mu}\right).
\end{eqnarray}

At this point of the discussion it should be noted that there are other sources of potential 
axions (axion meaning here, a field associated with a Peccei Quinn symmetry) in string theories than the universal one. These are arising from zero
modes of antisymmetric tensor field $B_{\alpha\beta}$ which is present in the supergravity multiplet of
the 10 dimensional theory. Components $B_{ij}$ (where $i,j$ are indices tangent to the 
six-dimensional compact
space) can have axion-like couplings in a scheme dependent way (in contrast to the universal
axion which can be seen as coming from the components $B_{\mu\nu}$ tangent to the 4-dimensionnal
non-compact space). In type $I$ and $II$ theories the scalars from the $R-R$ sector are also potential
axions. For the universal axion, as well as for the others, the exact pattern of symmetry one ends up with in
the 4-dimensional theory is very scheme-dependent since these fields receive
mass from instanton effects (field theory, world-sheet, or Dirichlet instantons) breaking the symmetry
they are associated with. The universal axion, e.g., receives mass contributions from instantons of all
the gauge groups which survive under the string scale, possibly including other groups than 
the standard model gauge groups.
We did not include mass terms for the axion in our lagrangian (\ref{susypot},\ref{lag}) since 
they are very model dependent, and
 are suppressed by temperature in the early Universe where cosmic strings are likely to form 
 (they can
 however have important effect when the temperature of the universe is decreasing). Moreover, one
  can look at the lagrangian (\ref{lag}) as the most general one for an axion compensating some 
  anomalous $U(1)_{X}$ symmetry, the couplings of the axion to the gauge fields in (\ref{lag}) being
  imposed by gauge invariance. In this latter case $M_{_P}$ has to be understood as the mass scale
  associated with the relevant theory.

\section{Higgs mechanism}

Let us now work out the Higgs mechanism in
this context. We consider for the sake of simplicity a single scalar
field $\Phi$ of negative charge $X$ and we drop consequently the $i$
indices.  ${\cal L}_{kin,\theta}$ can be rewritten

\begin{eqnarray}
\lefteqn{{\cal L}_{kin,\theta}=-\left[ M_{_P}^2\delta^2_{_{GS}} 
+ \phi^{2}X^{2}\right]\times}\nonumber
\\
& &\left[A^\mu-\frac
 {\frac{1}{2}M_{_P}\delta_{_{GS}}\partial^\mu a+\phi^2X\partial^\mu
\eta}{M_{_P}^2\delta_{_{GS}}^2+\phi^2X^2}\right]^2\nonumber\\
&& -\frac{\phi^2M_{_P}^2\delta^2_{_{GS}}X^2}{
M_{_P}^2\delta_{_{GS}}^2 + \phi^2X^2}\left[\frac
{\partial^{\mu}a}{2M_{_P}\delta_{_{GS}}}-\frac
{\partial^\mu\eta}{X}\right]^2 \nonumber\\
&& +\frac{\delta_{_{GS}}}{2g^2}\left(\frac{\phi^2X^2}{M_{_P}^2
\delta_{_{GS}}^2+\phi^2X^2}\left[\frac{a}{2M_{_P}\delta_{_{GS}}}
-\frac{\eta}{X}\right]\right.
\nonumber\\
&&\left.+ \frac {\frac{1}{2}M_{_P}\delta_{_{GS}}a+\phi^2X \eta}
{M_{_P}^2\delta_{_{GS}}^2+\phi^2X^2}\right) F_{\mu\nu}\widetilde
F^{\mu\nu}
-\frac{1}{4g^2}F_{\mu\nu}F^{\mu\nu} .
\end{eqnarray}

The linear combination appearing in this last equation
\begin{equation}
\frac{a}{2M_{_P}\delta_{_{GS}}}-\frac{\eta}{X}
\end{equation}
is the only gauge invariant linear combination of $\eta$ and $a$ (up
to a constant).  The other one
\begin{equation}
\ell \equiv \frac {\frac{1}{2}M_{_P}\delta_{_{GS}}a+\phi^2X \eta}
{M_{_P}^2\delta_{_{GS}}^2+\phi^2X^2}
\end{equation}
has the property of being linearly independent of the previous one and
of transforming under a gauge transformation ~(\ref{gauge}) as $\ell
\rightarrow \ell + \alpha$.  We now assume explicitely that $\Phi$
takes its vacuum expectation value $\langle \Phi^\dagger \Phi
\rangle \equiv \rho^2$ in order to minimize the potential~(\ref{pot}):
\begin{equation}
\rho^2=-\delta_{_{GS}}M_{_P}^2/X.\label{12}
\end{equation}
We are left, among other fields, with a massive scalar Higgs field
corresponding to the modulus of $\Phi$ of mass $m_{_X}$ given by
\begin{equation}
m_{_X}^2 = 2g^2\rho^2X^2=-2\delta_{_{GS}}Xg^2M_{_P}^2
\end{equation}
and we define
\begin{equation}
\hat a \equiv \left[\frac{a}{2M_{_P}\delta_{_{GS}}}-
\frac{\eta}{X}\right]\frac{\sqrt{2}\rho M_{_P}\delta_{_{GS}}X}
{\sqrt{M_{_P}^2\delta_{_{GS}}^2+\rho^2 X^2}}\label{hata}
\end{equation}
and 
\begin{eqnarray}
F_{a}^2 &= &\frac{1}{128\pi^4}\frac{M_{_P}^2 g^{4}}{\rho^2 X^2}
\left( M_{_P}^2 \delta^2_{_{GS}}+\rho^2X^2\right) \nonumber\\
 &=&\frac{1}{128\pi^4}M_{_P}^2 g^4
\left(1+\left(\frac{m_{_X}}{M_{_P}}\right)^2\frac{1}{2g^2X^2}\right)
\end{eqnarray}
so that with $\rho$ being set:
\begin{eqnarray}
\lefteqn{{\cal L}_{kin,\theta} = \nonumber} \\
&& -\left[ M_{_P}^2\delta_{_{GS}}^2
+\rho^2X^2\right]\left[A^{\mu} -\partial ^\mu \ell\right]^2
-\frac{1}{2}\partial^{\mu}\hat a \partial_{\mu} \hat a \nonumber \\
&&+\left[\frac{\hat
a}{32\pi^2F_{a}}+\frac{\delta_{_{GS}}}{2g^2}\ell\right]F_{\mu\nu}
\widetilde F^{\mu\nu}-\frac{1}{4g^2}F_{\mu\nu}F^{\mu\nu}
\end{eqnarray}
we can now make a gauge transformation to cancel $\partial ^\mu \ell $
by setting $\alpha = -\ell $.  This leaves us with
\begin{eqnarray}
\lefteqn{{\cal L}_{kin,\theta}= -\frac{m^2_{_A}}{2g^2}A^{\mu}
A_{\mu}-\frac{1}{2}\partial^{\mu}\hat a \partial_{\mu} \hat a
}\nonumber\\ &&+\frac{\hat a}{32\pi^2F_{a}}F_{\mu\nu}
\widetilde F^{\mu\nu}-\frac{1}{4g^2}F_{\mu\nu}F^{\mu\nu} ,
\end{eqnarray} 
where $m_{_A}$ given by
\begin{eqnarray}
m_{_A}^2 &=& 2g^2\left[\rho^2X^2+M^2_{_P}\delta_{_{GS}}^2\right]\nonumber \\
&= &m_{_X}^2 \left[
1+\left(\frac{m_{_X}}{M_{_P}}\right)^2\frac{1}{2g^2X^2}
\right] \label{mA}
\end{eqnarray}
is the mass of the gauge field after the symmetry breaking.
The remaining symmetry
\begin{equation}
\hat a \rightarrow 
\hat a  + \frac {32\pi^2F_a}{2g^2}\delta_{_{GS}}\beta
\end{equation}
is the rigid Peccei-Quinn symmetry which compensates for the anomalous
term arising from a rigid phase transformation of parameter $\beta$ on the fermions.

To summarize we have seen that in the presence of the axion the gauge
boson of the pseudo-anomalous symmetry absorbs a linear combination
$\ell$ of the axion and of the phase of the Higgs field. We are left
with a rigid Peccei-Quinn symmetry, the remnant axion being the other
linear combination $\hat a$ of the original string axion and of the
phase of the Higgs field.

\section{Pseudo-anomalous U(1) strings}
\subsection{cosmic string solutions}

We now look for stationary local cosmic string solutions of
 the field equations derived from Eq.~(\ref{lag}) provided the 
underlying U(1) symmetry
is indeed broken, which implies that at least one of the eigenvalues
$X_i$ is negative.  So we shall in this section assume again only one field $\Phi$ with
charge $X$, with $X<0$.
The fields equations  are:

\begin{equation} \Box a = 2\delta_{_{GS}} M_{_P} \partial_\mu
A^\mu -{1\over 2g^2 M_{_P}}F_{\mu\nu}\widetilde F^{\mu\nu},
\label{Boxa}\end{equation}
\begin{equation} \Box \phi = \phi (\partial _\mu\eta-XA_\mu)^2 
+g^2X\phi (X\phi^2 +\delta_{_{GS}} 
M_{_P}^2),\label{Boxphi}\end{equation}
\begin{equation} \partial_\mu [\phi^2 (\partial^\mu \eta - X A^\mu)]
= 0,\label{Boxeta}\end{equation}
\begin{eqnarray}{1\over g^2} \partial_\mu ({a\over M_{_P}}\widetilde 
F^{\mu\nu} - F^{\mu\nu})&=&\delta_{_{GS}} M_{_P} \partial ^\nu a
-2\delta_{_{GS}}^2 M_{_P}^2 A^\nu \nonumber\\ & & +2X\phi^2
(\partial^\nu\eta-XA^\nu),\label{BoxA}\end{eqnarray}
Using  Eq.~(\ref{deftilde}), which implies $\partial_\mu \widetilde
F^{\mu\nu} =0$, and deriving Eq.~(\ref{BoxA}) with respect to
$x^\nu$. We obtain with the help of Eqs.~(\ref{Boxa}) and (\ref{Boxeta}),
\begin{equation} F_{\mu\nu}\widetilde F^{\mu\nu} = 0,\label{FFa}\end{equation}
and we can rewrite Eq.~(\ref{BoxA}),
\begin{equation} {1\over g^2}\partial_\mu F^{\mu\nu} = {1\over M_{_P}}
\widetilde F^{\mu\nu}\partial_\mu a + {\cal J}^\nu +
J^\nu,\label{BoxA2} \end{equation}
where the currents are defined as
\begin{equation} J^\mu = -2 X\phi^2 (\partial^\mu\eta -
XA^\mu) = -2X\phi {\cal D}^{\mu}\eta \label{J}, \end{equation}
and
\begin{equation}{\cal J}^\mu = -\delta_{_{GS}}  M_{_P}
(\partial ^\mu a-2\delta_{_{GS}} M_{_P}A^\mu)=-2M_{_P}\delta_{GS}{\cal D}^\mu a.
\label{calJ}\end{equation}
Eqs.~(\ref{Boxa}) and (\ref{Boxeta}) then simply express those two
currents conservation $\partial \cdot J = \partial \cdot {\cal J} =0$,
when account is taken of Eq.~(\ref{FFa}).

Looking for a local cosmic string solution, we ask that on a 1-circle at infinity on a 2-plane
transverse to the string, the kinetic and potential energy of the different fields 
vanish. Namely:

\begin{eqnarray}
F^{\mu\nu}F_{\mu\nu}=0, \label{cinjauge}\\
(D^\mu \Phi)^2 = 0,  \label{cinphi}\\
({\cal D}^\mu a)^2 = 0,  \label{cina}\\
V(\Phi)=0. \label{potphi} 
\end{eqnarray}
As usual the fact that the $\Pi_1$ of the vacuum manifold derived from the potential (\ref{pot}) is non trivial
leads to the possibility to have non trivial winding solution of the equation (\ref{potphi}) with an 
asymptotic behaviour is in cylindrical coordinates,
\begin{eqnarray} \Phi = \rho \hbox{e}^{i\eta}, \\
\eta= n \theta, \end{eqnarray}
where $\rho$ is defined in (\ref{12}) and $n$ is the string winding number . Equation (\ref{cinphi}) and 
(\ref{cinjauge}) can then be satisfied asymptotically by taking $A_{\mu}$ a pure gauge 
and in such a way as to compensate for the Higgs field
energy density:
\begin{equation}
A_\mu = \partial_\mu\eta/X.
\end{equation}
 as in the Nielsen-Olesen \cite{NO} solution. And
equation (\ref{cina}) induces then a winding of the axion field with a winding number related to that
of $\eta$ by 

\begin{equation} a = {2 \delta_{_{GS}}M_{_P} \over X} \eta,\label{a}
\end{equation}
a perfectly legitimate choice as it should be remembered that $a$
is a periodic field of period $4\pi
\delta_{_{GS}}M_{_P}$ ($\mid X \mid$ is here implicitely equal to $1$, but it is clear that
for any other value a solution such as (\ref{a}) exists) . 

The energy of the cosmic string configuration is confined in the string like in the 
Nielsen-Olesen strings (and for the very same reason) and the cosmic string is perfectly local. This is in striking
contrast with the case of a global string where a divergent behavior of the energy density 
arises because the energy is not localized and a large distance
cut-off must be introduced. In this case, a divergence is still to be found as we will now see, but this
time at a small distance near the string core so that the total
energy is localised in a finite region of space.

The stress energy tensor is given by:
\begin{equation} T^\mu_\nu = -2g^{\mu\gamma} {\delta {\cal L}\over
\delta g^{\gamma\nu}}+\delta^\mu_\nu {\cal L},\end{equation}
which reads explicitely
\begin{eqnarray} T^{\mu\nu}&=&2[\partial^\mu\phi\partial^\nu\phi
-{1\over 2}g^{\mu\nu}(\partial \phi)^2]\nonumber \\ & & +{1\over g^2}
(F^{\rho\mu} F_\rho ^{\ \nu}-{1\over 4}g^{\mu\nu}F\cdot F)
\\
& & -{1\over 2} g^2 g^{\mu\nu} (X\phi^2+
\delta_{_{GS}}M_{_P}^2 )^2\nonumber\\
& & +{1\over 2\delta_{_{GS}}^2 M_{_P}^2 } [{\cal J}^\mu {\cal
J}^\nu -{1\over 2} g^{\mu\nu} {\cal J}^2]\nonumber\\ & & +{1\over 2
X^2
\phi^2} [J^\mu J^\nu -{1\over 2} g^{\mu\nu} J^2]
\nonumber\\
\label{Tmunu}\end{eqnarray}
where account has been taken of the field equations. The energy per
unit length $U$ and tension $T$ will then be defined respectively as
\begin{equation} 
U=\int d\theta\, rdr T^{tt} \ \ \hbox{ and }\ \ 
T=-\int d\theta\,rdr T^{zz},\label{UT}
\end{equation}
The question as to whether the corresponding string solution is local
or global is then equivalent to asking whether these quantities are
asymptotically convergent ({\em i.e.} at large distances).
The total energy per unit length (and tension) is however not finite
in this simple string model for it contains the term:
\begin{equation} U = \hbox{f.p.} + 2 \pi \int {dr\over r}
({\delta_{_{GS}} M_{_P} n\over X} - \delta_{_{GS}} M_{_P} A_\theta
)^2,\end{equation} (f.p. denoting the finite part of the integral) so
that, since $A_\theta$ must vanish by symmetry in the string core, one
ends up with
\begin{equation} U = \hbox{f.p.} + 2 \pi  ({\delta_{_{GS}}
M_{_P} n\over X} )^2 \ln ({R_A\over r_a}),\label{div}\end{equation}
where $R_A$ is the radius at which $A_\mu$ reaches its asymptotic
behaviour, i.e., roughly its Compton wavelength $m_{_{A}}$ given in
(\ref{mA}), while $r_a$ is defined as the radius at which the
effective field theory (\ref{lag}) ceases to be valid, presumably of
order $M_{_P}^{-1}$; the correction factor is thus expected to be of
order unity for most theories. Hence, as claimed, the strings in this
model can be made local with a logarithmically divergent energy. The
regularization scale $r_a$ is however a short distance cut-off, solely
dependent on the microscopic structure and does involve neither
the interstring distance nor the string curvature radius. In particular, the
gravitational properties of the corresponding strings are those of a
usual Kibble-Vilenkin string~\cite{vilenkin}, given the equation of
state is that of the Goto-Nambu string $U=T=$const., and the light
deflection is independent of the impact parameter~\cite{defl}.

The solution (\ref{a}) turns out, as can be explicitely checked using
Eqs.~(\ref{Boxa}) and (\ref{FFa}), to be the only possible non trivial
and asymptotically converging solution. 

Moreover, the stationnary solution
(\ref{a}) shows the axion gradient to be orthogonal to $\widetilde
F^{\mu\nu}$, i.e., $\partial_\mu a
\widetilde F^{\mu\nu}=0$. Therefore, Eqs.~(\ref{Boxa}-\ref{BoxA2})
reduce to the usual Nielsen-Olesen set of equations~\cite{NO}, with
the axion coupling using the string solution as a source term. It is
therefore not surprising that the resulting string turns out to be
local.

The local string solution we have found can also be considered using the 
new dynamical variables $\hat a$ and $\ell$ defined in the previous section. 
In this langage the local strings considered are the one obtained by a winding of $\ell$ and
$A_\mu$ around the string core, whereas $\hat a$ is not winding. A winding of $\hat a$ would
generate global axionic-like strings decoupled from the previous ones.

\subsection{Local string genesis}

Forming cosmic strings during a phase transition is a very complicated
problem involving thermal and quantum phase
fluctuations~\cite{zurek}. As it is far from being clear how  $a$
and $\eta$ fluctuations will be correlated (even though they presumably
will be). 
One can consider, as a toy model, the possibility that a network
of two different kinds of strings will be formed right after the phase
transition, call them $a-$strings and $\eta-$strings, with the meaning
that an $a-$string is generated whenever the axion field winds
(ordinary axion string) while an $\eta-$string appears when the Higgs
field $\Phi$ winds. Both kinds of strings are initially global since
for both of them, only part of the covariant derivatives can be made
to vanish.
We however expect the string network to consist, after some time, in
only these local strings together with the usual global axionic
strings.

Let us consider an axionic string with no Higgs winding: as
$A^\mu\not= 0$, the vacuum solution $\Phi =\rho$ [Eq.~(\ref{12})] is
not a solution, and thus the axionic string field configuration is
unstable. As a result of Eq.~(\ref{Boxphi}), the Higgs field amplitude
tends to vanish in the string core. At this point, it becomes, near
the core, topologically possible for its phase to start winding around
the string, which it will do since this minimizes the total energy
while satisfying the topological requirement that $A_\mu$ flux be
quantized. Such a winding will propagate away from the string.

Conversely, consider the stability of an $\eta-$string with $a=0$. The
conservation of ${\cal J}$ implies, as one can fix $\partial_\mu
A^\mu=0$, that $\Box a =0$, whose general time-dependent solution is
$a = a(|{\bf r}|\pm t)$. Given the cylindrical symmetry, this solution
can be further separated into $a=f(r-t) \theta$. This means that
having a winding of $a$ that sets up propagating away from the string
is among the solutions. As this configuration ultimately would
minimize the total energy, provided $\lim
_{t\to\infty}f=-2 \delta_{_{GS}}M_{_P}/X$, this means that the original
string is again unstable and will evolve into the stationary solution
that we derived in the previous section. 

It should be remarked at this point that these time evolution can in
fact only be accelerated when one takes into account the coupling
between $a$ and $\eta$: if any one of them is winding, then the other
one will exhibit a tendency to also wind, in order to locally minimize
the energy density. Indeed, it is not even really clear whether the
string configurations we started with would even be present at the
string forming phase transition. What is clear, however, is that after
some time, all the string network would consist of local strings
having no long distance interactions. This means in particular that
the relevant scale, if no inflationary period is to occur after the
string formations, should not exceed the GUT scale in order to avoid
cosmological contradictions.

\subsection{Constraints on the scale of the symmetry breaking}
The cosmological evolution of the network of strings formed in these
theories may also lead to serious constraints on the Green-Schwarz
coefficient. If domain walls form connecting the strings, which itself depends on the
(temperature-dependent) potential generated by instantons,
 the network is known to rapidely (i.e. in less than a
Hubble time) decay into massive radiation and the usual constraint
relative to the axion mass would hold~\cite{kibble,wall-string}. If
however the string network is considered essentially stable, then its
impact on the microwave background limits the symmetry breaking scale
$\delta _{_{GS}} M_{_{P}}$ through the observational requirement that
the temperature fluctuations be not too large~\cite{dT-T}, i.e.  $$GU
\alt 10^{-6},$$ with $G$ the Newton constant $G=
M_{_{P}}^{-2}/(8\pi^2)$. Therefore, the cosmological constraint reads
\begin{equation}
\delta_{_{GS}}  \alt 10^{-2}, \label{delcons}
\end{equation}
a very restrictive constraint indeed which can be compared with the scale given 
in (\ref{weaks}) or with similar predictions in the strongly coupled heterotic strings \cite{nous2}.

The strings that we have discussed here might appear in connection
with a scenario of inflation. Indeed, the potential (\ref{pot}) is
used for inflation in the scenario known as $D$-term
inflation~\cite{Dterm}: inflation takes place in a direction neutral
under $U(1)$ and the corresponding vacuum energy is simply given by:
\begin{equation}
V_0 = {1 \over 2} g^2 \delta^2_{_{GS}} M_{_P}^4.
\end{equation}
The $U(1)$-breaking minimum is reached after inflation, which leads to
cosmic strings formation. Such an inflation era cannot therefore
dilute the density of cosmic strings and one must study a mixed
scenario~\cite{jeannerot}.  It is interesting to note that, under the
assumption that microwave background anisotropies are predominantly
produced by inflation, the experimental data puts a
constraint~\cite{ERR,LR,KM-R} on the scale $\xi \equiv
\delta_{_{GS}}^{1/2} M_{_P}$ which is stronger than
(\ref{delcons}). Several ways have been proposed~\cite{ERR,KM-R} in
order to lower this scale. They would at the same time ease the
constraint (\ref{delcons}).

\subsection{string currents}

It is clear that the model (\ref{lag}) is no longer supersymmetric, since we have set the dilaton
to its v.e.v, whereas the axion is a dynamical field and belongs to the same multiplet.
 Allowing the dilaton to be dynamic as in (\ref{susypot}), we can easily preserve supersymmetry after the breaking
 of the $U(1)$; which can be desirable if we assume that supersymmetry is broken at some lower scale than the
 one of $U(1)$ breaking. The $D$-term which gives the potential (\ref{pot}) being zero after 
 symmetry breaking, one has only to be sure that the v.e.v taken by the Higgs field $\Phi$ does not
 destabilize the vanishing energy of the vacuum through the superpotential (an explicit example in given
 in the first reference of {\cite{U1}}).  
  Following Hughes and Polchinski \cite{hupo} we expect
  that, even when the lagrangian is fully supersymmetric, at least
some of the supersymmetry generators will be broken by the cosmic string configuration because this
configuration breaks translational invariance.  This will lead to
goldstinos wich are massless Fermi fields on the string arising from Fermi zero modes in the
underlying theory and give rise to supercurrents \cite{witten} (some explicit example of these
superconducting 
supersymmetric cosmic strings have recently been worken out by S.C. Davis, A.C. Davis and M.Trodden
\cite{sdad,susystri}). These currents which can also appear in a non supersymmetric model from the
coupling  of the string to other fields, fermionic in particular, tend to raise the
stress-energy tensor degeneracy in such a way that the energy per unit
length and tension become dynamical variables. For loop solutions,
this means a whole new class of equilibrium solutions, named vortons,
whose stability would imply a cosmological
catastrophe~\cite{vortons}. If these objects were to form,
Eq.~(\ref{delcons}) would change into a drastically stronger
constraint. Issues such as the explicit construction of the currents, their relation to
supersymmetry   and  whether supersymmetry breaking might
destabilize the currents~\cite{sdad}, thereby effectively curing the
model from the vorton problem, still deserve investigation; as do the possible consequences of
having a 
 dynamical dilaton \cite{damou}. We are also planing to look at a more
realistic model in the framework of horizontal symmetries.

\section*{acknowledgement}
I wish to thank the organizers of the third Peyresq meeting on cosmology which was so
stimulating and interesting, especially Mrs. Smet for providing such a warm and
welcome atmosphere during the meeting. I am also very gratefull  to Pr. Edgar Gunzig
 and Brandon Carter for having invited me, and to Mrs. Gunzig for her hospitality.

\end{document}